\documentclass[twocolumn,showpacs,preprintnumbers,amsmath,amssymb]{revtex4-1}

\usepackage{graphicx}
\usepackage{dcolumn}
\usepackage{amsfonts,amsmath,amssymb,bm}

\begin{document}

\title{\bf {Electrical Transport Properties in ZnO Bulk, c/ZnO and ZnMgO/ZnO/ZnMgO Heterostructures}}
\date{\today} 
\author{M. Amirabbasi$^a$}
\author{I.~Abdolhosseini~Sarsari$^a$}
\affiliation{a) Department of Physics, Isfahan University of Technology, Isfahan, Iran}
\begin{abstract}
In this paper, the reported experimental data in [Sci. Rep., 2012, 2, 533] related to electrical transport
properties in bulk ZnO, ZnMgO/ZnO, and ZnMgO/ZnO/ZnMgO single and double 
heterostructures were analyzed quantitatively and the most important 
scattering parameters for controlling electron concentration and electron 
mobility were obtained. 
Treatment of intrinsic mechanisms included polar-optical phonon scattering, 
piezoelectric scattering and acoustic deformation potential scattering. 
For extrinsic mechanisms, ionized impurity, dislocation scattering, and 
strain-induced fields were included. 
For bulk ZnO, the reported experimental data were corrected for removing the 
effects of a degenerate layer at the ZnO/sapphire interface via a two layer 
Hall effect model.
Also, donor density, acceptor density and donor activation energy were determined
via the charge balance equation. This sample exhibited hopping conduction below 
50K and dislocation scattering closely controlled electron mobility closely.
The obtained results indicated that the enhancement of electron mobility 
in double sample, compared with the single one, can be attributed to the 
reduction of dislocation density, two dimensional impurity density in the 
potential well due to background impurities, and/or interface charge and 
strain-induced fields, which can be related to better electron confinement 
in the channel and enhancement in the sheet carrier concentration of 2DEG in this sample.  
\end{abstract}

\maketitle

\section{Introduction}
ZnO has received substantial interest of the research community due to its wide band gap
(3.4eV ~\cite{Dai2011}), high breakdown voltage, generating less noise, high temperature power, and 
sustaining large electrical field ~\cite{Meng,kpark,Pholnak, Morkoc}. ZnO is used in 
a variety of optical and optoelectronic applications such as UV light – emitting diodes ~\cite{Wang,Wnag2}, 
transparent transistors~\cite{Kwon}, UV detectors ~\cite{Guoa}, and UV laser diodes ~\cite{Shi,Szymanski}. 
In bulk ZnO, mobility decreases at low and high temperatures due to dislocation scattering,
ionized impurity, and lattice vibrations, respectively. Undoped ZnO with a wurtzite structure is naturally
an n-type semiconductor due to the presence of intrinsic defects such as Zn interstitial and O
vacancy~\cite{Morkoc}. To make high quality P-type ZnO, knowing native defects in an undoped ZnO 
via Hall-effect measurements is essential. To obtain high mobility especially at low temperatures, 
ZnMgO/ZnO heterostructures are used ~\cite{Bian,Sang}, in which electrons see a two-dimensional space.
Consequently, a two – dimensional electron gas (2DEG) is formed at the interface due to the internal
electric field. 
Since carrier confinement can influence electron mobility, ZnMgO/ZnO/ZnMgO double quantum well has been
fabricated recently ~\cite{Wang3,Kuznetsov}. The ability to fabricate a ZnMgO-heterostructure makes 
the fabrication of ZnMgO-high electron mobility transistors (HEMT) possible~~\cite{Barquinha}, which 
have received more attention recently ~\cite{Amirabbasi,Ji,Meng3}. Transport properties such as
carrier mobility ($\mu$) and
carrier concentration (n) are crucially important, because the operation of all these devices depends critically
on current transport. 

In this paper, the reported experimental data related to electrical transport properties
in bulk ZnO, ZnMgO/ZnO and ZnMgO/ZnO/ZnMgO single and double heterostructures were analyzed quantitatively,
which is reported by J. Ye et al ~\cite{Ye}. The bulk $ZnO, Zn_{0.82}Mg_{0.18}O$ (75nm) /ZnO single (sample A), 
and $Zn_{0.8}Mg_{0.2}O(60nm)/ZnO(30nm)/graded-Zn_{0.85}Mg_{0.15}O(90nm)$ double (sample B)
heterostructures were grown on sapphire using metal-organic vapor phase epitaxy technique ~\cite{Ye}.
The experimental details were given in ~\cite{Ye}. Also, the sheet carrier concentrations (n$_s$) 
of the 2DEG in samples A, and B were reported as 1.48$\times$ $10^{12}~cm^{-2}$ and
1.16 $\times$ $10^{14}~cm^{-2}$, respectively ~\cite{Ye}.

\section{Theory of Electrical properties}
\subsection{Charge balance equation}

Charge balance equation was used for the carrier concentration data in which semiconductor was
assumed to be n-type and non-degenerate ~\cite{Look}:
\begin{equation}
n+N_\mathrm{a}=\sum \frac {N_\mathrm{di}}{1+n/\phi_\mathrm{i}}
\end{equation}
where 
\begin{equation}
\phi_\mathrm{i}=\frac {g_\mathrm{0i}}{g_\mathrm{i}}N_\mathrm{c}exp(\frac {\alpha_\mathrm{i}}{k_\mathrm{B}})exp(\frac
{-E_\mathrm{d0i}}{k_\mathrm{B}T})
\end{equation}
\begin{equation}
N_\mathrm{c}=2(2 \pi m^* k_\mathrm{B} /h^2)^{3/2}
\end{equation}
in which, $N_\mathrm{a}$ is acceptor density, $N_\mathrm{d}$ is donor density,
$g_\mathrm{0i}$ ($g_\mathrm{i}$) is the unoccupied (occupied) state degeneracy of
donor i, and $\alpha$ is the temperature coefficient defined by 
$E_\mathrm{di}=E_\mathrm{0i}-\alpha_\mathrm{i} T$ 
in which $E_\mathrm{d}$, and $E_\mathrm{d0}$ are the activation energy of the donor electrons at T 
and zero temperature, respectively ~\cite{Morkoc}
($\alpha$ is assumed zero ~\cite{Morkoc}). 
It should be noted that $N_\mathrm{a}$, $N_\mathrm{d}$, and $E_\mathrm{d}$ are considered 
fitting parameters.

\section{Theory of scattering mechanisms in bulk semiconductors} 

Here, mobility limitation due to each individual scattering process is calculated independently, 
using the corresponding analytical expressions ($\mu_\mathrm{i}$). The total mobility can be calculated from
the scattering limiting motilities, using Matthiessen's rule 
(${1}/{\mu_\mathrm{tot}} = \sum {1}/{\mu_\mathrm{i}}$).
The material parameters used in the calculations are listed in~\ref{tab1}
\begin{table}[htb]
\caption{\label{tab1}~ZnO material parameters used in the calculations.}
\begin{ruledtabular}
\begin{tabular}{l|l} 
Material parameters &Values\\
Density of crystal $(\rho)$~\cite{Furno}                                   &$6.1 \times 10^{3} (kgm^{-3})$  \\
Deformation potential energy $(E_{L})$~\cite{Rode}                         &$3.5 (eV)$                      \\
High-frequency dielectric constant $(\varepsilon_{\infty})$~\cite{Look2}   &$3.72\varepsilon_{0} (Fm^{-1})$ \\
Static dielectric constant $(\varepsilon_{s})$~\cite{Look2}                &$8.12 \varepsilon_{0} (Fm^{-1})$\\
Effective mass $(m*)$ ~\cite{Look2}                                        &$0.3m0 (kg)$                    \\
ZnO lattice constant $(a_{0})$~\cite{Amirabbasi}                           &$0.521(nm))$                    \\
Piezoelectric constant $(h_{pz})$~\cite{Furno}                             &$1.10(Cm^{-2})$\\
Sound velocity $(s)$~\cite{Furno}                                          &$6.59 \times 10^{3} (ms^{-1})$\\
\end{tabular}
\end{ruledtabular}
\end{table}.

\subsection{Intrinsic scattering mechanisms}
(I)~Mobility ($\mu_{ac}$) of bulk electron being scattered from acoustic deformation potential scattering, is
given by~\cite{Anderson}:

\begin{equation}
\mu_\mathrm{ac}(T)=\frac{2(2\rho)^{0.5}\rho s^{2} \hbar^{4} e}  {3 E_\mathrm{L}^{2} (m^{*})^{2.5} 
(k_\mathrm{B}T)^{1.5}}
\end{equation}

where $E_{L}$, $\rho$, and s are deformation potential energy, density of the crystal, and sound velocity,
respectively.

(II)~Mobility ($\rho_{pz}$) limited by piezoelectric scattering is expressed as~\cite{Anderson}:

\begin{equation}
\mu_\mathrm{pz}(T)=\frac{16(2\pi)^{0.5}\rho s^{2} \hbar^{2} e}  {3 (eh_\mathrm{pz} /\varepsilon_\mathrm{s} 
\varepsilon _\mathrm{0})^{2} (m^{*})^{1.5} (k_\mathrm{B}T)^{0.5}}
\end{equation}
in which, $h_\mathrm{pz}$ is the piezoelectric constant in ZnO.

(III)~Mobility ($\mu_{pop}$) caused by polar optical phonon scattering, which controls carrier mobility at high
temperature, can be calculated using~\cite{Ehrenreich}:
 
\begin{equation}
\mu_\mathrm{pop}(T)=0.199(T/300)^{0.5} (e/e^{*})^{2} (m/m^{*})^{1.5} 10^{22}~M 10^{23}(A)
\end{equation}
$A=\frac{\hbar\omega}{k_\mathrm{B}(T) -1}  G(\frac{\hbar \omega}{k_\mathrm{B}(T)})10^{-13} V_\mathrm{a}\omega$, 

Here, $\omega$, $e^{*}=(M V_{a} \omega \varepsilon_{0} (1/\varepsilon_{\infty} - 1/\varepsilon_{s}))^{0.5}$, $V_{a}$,
and M are polar phonon frequency, Callen's effective ionic charge, volume for a Zn and O ion
pair, and reduced ionic mass, respectively.

\subsection{Extrinsic scattering mechanisms}
(I) Mobility ($\mu_{Im}$) determined by ionized impurity scattering is given by~\cite{Brooks}:
\begin{equation}
\mu_\mathrm{Im}(T)=\frac{128(2\pi)^{1/2}(\varepsilon_{s}\varepsilon_{0})^{2}(k_{B}T)^{3/2}[ln(1+b)-b/b+1]^{-1}}{e^{3}
((m^{*})^{0.5})(n+2N_{a})}
\end{equation}
where, N$_a$ is acceptor density obtained from equation (1) as the fitting parameter.

(II)~The mobility ($\mu_{cd}$) caused by crystalline defects (domain boundaries and strain induced fields) 
scattering can be obtained from~\cite{Tang}:

\begin{equation}
\mu_\mathrm{cd}(T)=\frac{C}{T^{1.5}}
\end{equation}
Here, C is attributed to the strained induced fields and domain boundaries.

(III)~Mobility limited ($\mu_{disl}$) by dislocation scattering which controls carrier 
mobility at low temperature, is expressed as~\cite{Podor}: 

\begin{equation}
\mu_\mathrm{disl}(T)=\frac{30(2\pi)^{0.5}(\varepsilon_{0}\varepsilon_{s})
a^{2}(k_{B}T)^{1.5}}{N_{disl}e^{3}f^{2}\lambda_{D}(m^{*})^{0.5}}
\end{equation}
in which, $\lambda_{D}=(\frac{\varepsilon_{s}\varepsilon_{0}k_{B}T}{e^{2}n})^{0.5}$, $N_{disl}$, f (=1), and a
are dislocation density, 
occupancy rate, and distance between acceptor centers, respectively. 

\section{Two – layer Hall – effect model}

Due to lattice mismatch between semiconductor layer and substrate, a narrow area with high dislocation density
(degenerated layer) is formed at semiconductor layer/ substrate interface and has crucial effects on
the electrical properties (carrier concentration and mobility) of semiconductor especially in low temperatures.
We can use two-layer Hall-effect model to correct the carrier concentration and mobility experimental 
data~\cite{Look5}:

\begin{equation}
n_\mathrm{1}=\frac{(\mu_{H} n_{H}-\mu_{2} n_{2})^{2}}{(\mu_{H})^{2} n_{H}-(\mu_{2})^{2} n_{2}}
\end{equation}
\begin{equation}
\mu_\mathrm{1}=\frac{(\mu_{H})^{2} n_{H} - (\mu_{2})^{2} n_{2}}{\mu_{H} n_{H}-\mu_{2}n_{2}}
\end{equation}
where $\mu_{H}$, and $n_{H}$ are experimental mobility and carrier concentration and also 
$\mu_{2}$, and $n_{2}$  are  mobility and carrier concentration in degenerated layer; 
$\mu_{1}$, and $n_{1}$ are the corrected data.

\section{Theory of scattering mechanisms in 2DEG}

Different scattering mechanisms are considered to model 2DEG mobility using the Matthiessen's rule. 
The analytical expressions of scattering mechanisms for 2DEG mobility are briefly summarized below 
and the relevant material parameters are listed in Table1.

\subsection{Intrinsic scattering mechanisms}

(I)~Polar optical phonon scattering is expressed as~\cite{Ridley,Hirakawa}:
\begin{equation}
\mu_\mathrm{pop}(T)=\frac{4\pi\varepsilon_{0}\varepsilon_{p}}{e\omega ((m)^{*})^{2} L} exp((\hbar\omega/k_{B}T)-1)
\end{equation}
where, L (2$(\frac{n_s}{10^{12}cm^{-2}})^{-1/3}\times(55\AA{})$) ~\cite{Lee} is the width of quantum well and 
$1/\varepsilon_{p}=1/\varepsilon_{\infty}-1/\varepsilon_{s}$, 
n$_s$ is 2DEG sheet carrier density.

(II)~Acoustic deformation potential scattering is given by~\cite{Basu}:
     
\begin{equation}
\mu_\mathrm{ac}(T)=\frac{e\hbar^{3}\rho (u_{L})^{2}L}  {(m^{*})^{2} (E_{l})^{2} k_{B}T}
\end{equation}
Here, u$_l$ is the longitudinal acoustic phonon velocity.

(III)~In strongly polar materials, 
the most powerful interaction with acoustic phonons at low energies is via the piezoelectric effect. 
The piezoelectric scattering can be obtained from ~\cite{price}:

\begin{equation}
\mu_\mathrm{pz}(T)=\frac{\pi k_{f} (E_{l})^{2}}  {L e^{2} h_{14}^{2}} [9/32 + 13/32(u_{l}/u_{s})^{2} I_{A}(\gamma_{t}) /I_{A}
(\gamma_{l})]^{-1}
\end{equation}
where, $k_f (= (2πn_s)^ {1/2})$ is the wave vector on the Fermi surface, h$_{14}$ is the piezoelectric constant, 
u$_t$ is the transverse acoustic phonon velocity, and $I_{A}(\gamma_{t}(A))=((4\gamma_{t}/3)^{2} + 1)$, $I_{A}(\gamma_{l}(A))=
((4\gamma_{l}/3)^{2} + 1)$
$\gamma_{t}=\frac{2\hbar u_{t} k_{f}}{k_{B}(T)}$ ,$\gamma_{l}=\frac{2\hbar u_{l} k_{f}}{k_{B}(T)}$. 

\subsection{Extrinsic scattering mechanisms}

(I)~In heterostructures with 2DEG, although free electrons are separated from the ionized donors, 
they can still scatters from them. Mobility is caused by ionized impurity scattering due 
to the scattering of remote donors~\cite{Lee}:

\begin{equation}
\mu_\mathrm{remote}=\frac{64\pi \hbar^{3}(\varepsilon_{2})^2 (\varepsilon_{s})^{2}(S_{0})^{2} (k_{f})^3}  
{e\omega (m^{*})^{2}N_{d}} [1/(L_{0})^{2} - 1/L(M_{0})^{2}]{-1}
\end{equation}
where,
$L_{0}=d_{0} + L/2$ , $LM_{0}= L_{0}+d_{l}$
Here, d$_0$ is the width of the spacer layer, d$_l$ is the width of the depletion layer (n$_s$/N$_d$, with N$_d$ as the donor 
density in the barrier),
and S$_0$ ($=\frac{e^{2}m^{*}}{2\pi \varepsilon_{0}\varepsilon_{s}\hbar^{2}}$), the screening constant
~\cite{price1981}.

(II)~In ZnMgO/ZnO heterostructures, 2DEG is formed on the ZnO side of the ZnMgO heterointerface and hence background impurity
scatters free carriers, as well as due to interface charge; also:
Ionized impurity scattering due to interface 
charges can be calculated from ~\cite{Hess,Sah1972}:

\begin{equation}
\mu_\mathrm{remote}=\frac{4\pi \hbar^{3}(\varepsilon_{2})^2 
(\varepsilon_{s})^{2}(S_{0})^{2} (k_{f})^3}  {e^{3} (m^{*})^{2}N_{bi}I_{B}}
\end{equation}
where, N$_{bi}$ is the 2D impurity density in the potential well due to background impurities and/or interface charge and 
$I_{B}=\int \frac{sin\phi}{2sin\phi +(S_{0}/K_{f})^{2}}d\phi$

(III)~Dislocation scattering is expressed as ~\cite{Sarma,Davies}:

\begin{equation}
\mu_\mathrm{disl}=\frac{4\pi \hbar^{3}(\varepsilon_{2})^2 
(\varepsilon_{s})^{2}(S_{0})^{2} (k_{f})^4 c^{2}}  {e^{3} (m^{*})^{2}N_{disl}I_{t}}
\end{equation}
where, N$_{disl}$ is the charge dislocation density, c is the lattice constant of $In_{1-x}Al_{x}N [=xa_{l}(AlN)+(1-x)c(InN)$ 
suggested by Vegard's law],
$\xi$ is a dimensional parameter: $\xi=2k_{F}/qT_{F}; qT_{F}=2/a_{B}$ is the 2D Thomas-Fermi wave vector, where $a_{B} = 
\varepsilon_{s}
\varepsilon_{0}h^{2}/e^{2}\pi m^{*}$ 
is the effective Bohr radius in the material and $I_{t}=1/2\xi^{2}\int\frac{1}{(1+\xi^{2}u^{2})(1-u^2)^{0.5}}d\xi$

\section{Result and discussion}

\subsection{Bulk mode}

\begin{figure}
\includegraphics[width=0.40\textwidth,clip=true]{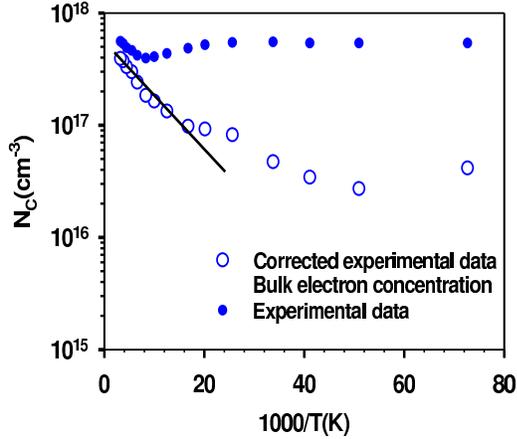}
\caption{\label{pslat1} Experimental and bulk (corrected) electron concentration 
versus temperature; the solid curve shows the fitting result to the corrected data.}
\end{figure}

\begin{figure}
\includegraphics[width=0.40\textwidth,clip=true]{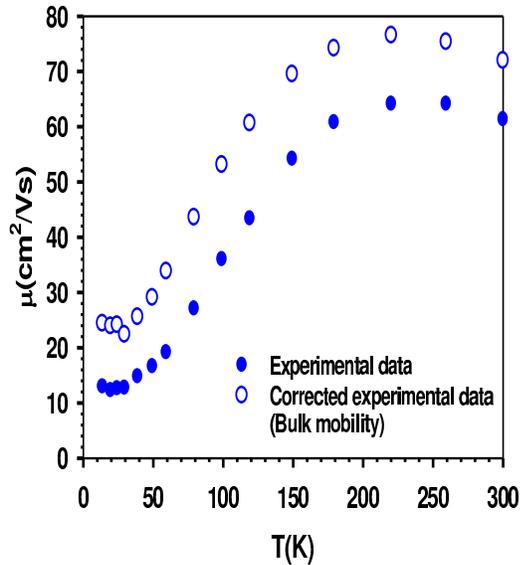}
\caption{\label{pslat2} Experimental and bulk (corrected) electron mobility versus temperature.}
\end{figure}

The experimental temperature – dependent electron concentration of ZnO is shown in Fig.~\ref{pslat1}. 
The experimental electron concentration decreases as temperature is decreased from 300 to 80 K, 
which is a carrier freeze out process, and then is increased slightly when temperature further
decreases. 
As a result, due to lattice mismatched between ZnO and sapphire, a two-dimensional parallel 
conduction layer (degenerate layer) is formed at the ZnO/sapphire interface, which is 
temperature-independent. For investigating the electrical transport properties of bulk layer, 
the electron concentration and electron mobility of degenerate layer should be removed via a 
two layer Hall effect model (see Eq. 10-11). The corrected data are 
shown in Fig.~\ref{pslat1} and Fig.~\ref{pslat2}.

As is clear in in Fig.~\ref{pslat1} and Fig.~\ref{pslat2}, by removing the effects of degenerate layer,
the bulk electron concentration reduces (5$\times$10$^{17}$ to 4.1$\times$10$^{16}$ cm$^{-3}$ at 15 K and 
5.22$\times$10$^{17}$ to 3.9$\times$10$^{17} cm^{-3}$ at~300K) and the bulk electron mobility
increases (15 to 20 $cm^{2}V^{-1}s^{-1}$ at 15K and 60 to 71 $cm^{2}V^{-1}s^{-1}$ at~300K) from 
their experimental values. For $T\textless50$ K, the slightly bulk electron concentration decrease 
indicates hopping conduction; then, good fitting between bulk data and theoretical curves is not obtained.
Temperature dependent bulk electron concentration is fitted by the charge balance 
equation (see Eq. 1). The obtained fitting values are listed in Table~\ref{tab2}. 
The obtained activation energy of residual donor is in reasonable agreement with
the reported value ~\cite{Look9}, which confirms the validity of the fitting.

\begin{figure}
\includegraphics[width=0.40\textwidth,clip=true]{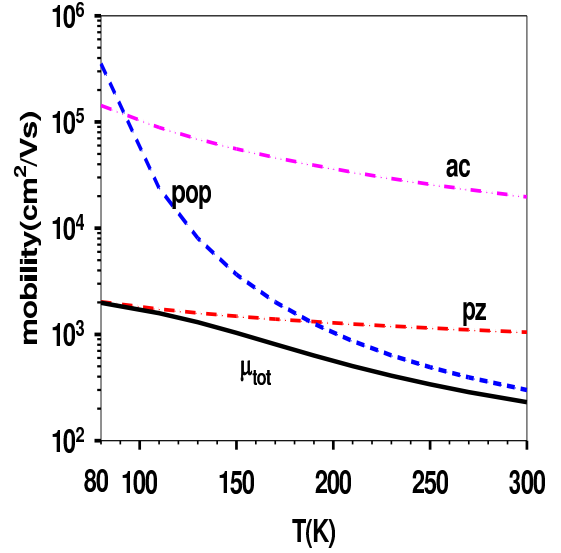}
\caption{\label{pslat3} Total mobility (solid curve) of hypothetically pure ZnO calculated 
by the Matthiesen's rule. The mobility limits due to the lattice scattering mechanisms are 
displayed in dash curves.}
\end{figure}

\begin{figure}
\includegraphics[width=0.40\textwidth,clip=true]{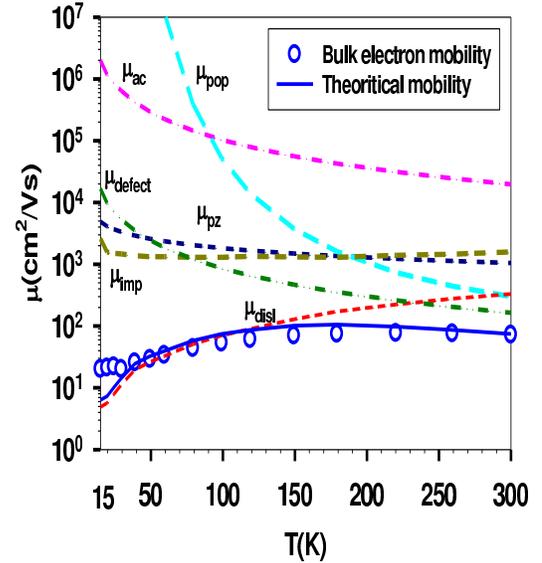}
\caption{\label{pslat4} The bulk electron mobility fittings for n-type ZnO sample via different 
scatering mechanisms.}
\end{figure}

Fig.~\ref{pslat3} shows the total mobility obtained from the Matthiesen's rule by considering the intrinsic scattering mechanisms. 
As seen in Fig.~\ref{pslat3}, in the pure ZnO sample, the total mobility at room temperature is in the order of 
10$^2$ ($cm^{2}V^{-1}s^{-1}$). 
Fig.~\ref{pslat4} shows the position of each scattering mechanisms for the bulk electron mobility of ZnO. The obtained fitting 
parameters are listed in Table~\ref{tab2}. It should be noted that N$_{disl}$ matches the accuracy of the reported values
~\cite{Vigue} in this regard. According to Fig.~\ref{pslat4}: 
In the low and medium temperature ranges, $15 \textless T \textless 240 K$, dislocation scattering is dominant.
In the high temperature range, $240 \textless T \textless 300 K$, polar optical phonon scattering controls 
the bulk electron mobility.

\begin{table}[htb]
\caption{\label{tab2} Value of the calculated fitting parameters for ZnO.}
\begin{ruledtabular}
\begin{tabular}{l|c} 
Fitting parameters                                        &Values\\
\hline
N$_a$  (cm$^{-3}$)                                        &$ 1 \times 10^{16} $\\
N$_d$  (cm$^{-3}$)                                        &$6 \times 10^{17}$\\
E$_d$  (meV)                                              &$25$\\
Dislocation density N$_{disl}$ (cm$^{-3}$)                &$6.5 \times 10^{13}$\\
Crystalline defects                                       &$8.5\times10^{5}$\\
\end{tabular}
\end{ruledtabular}
\end{table}

\subsection{2DEG mobility}

Regarding Vegard's law and the band gap values of ZnO (= 3.4 eV)~\cite{Dai2011} and MgO(= 5.88 eV), 
the band gap values of Zn$_{0.82}$Mg$_{0.18}$O/ZnO and Zn$_{0.8}$Mg$_{0.2}$O/ZnO/Zn$_{0.85}$Mg$_{0.15}$O
of about 3.88 eV and 3.89 eV, respectively, are calculated, which means the band gap of ZnMgO/ZnO is
wider than ZnO. 
So, it should be expected that the part of conduction electrons in the ZnMgO layer is transferred to
the adjacent layer with a smaller band gap (ZnO), which causes an internal field and subsequently the 
formation of a triangular quantum well; thus, the formation of a thin layer is resulted near the interface with a 
2DEG behavior~\cite{Baghani}.

\begin{figure}
\includegraphics[width=0.40\textwidth,clip=true]{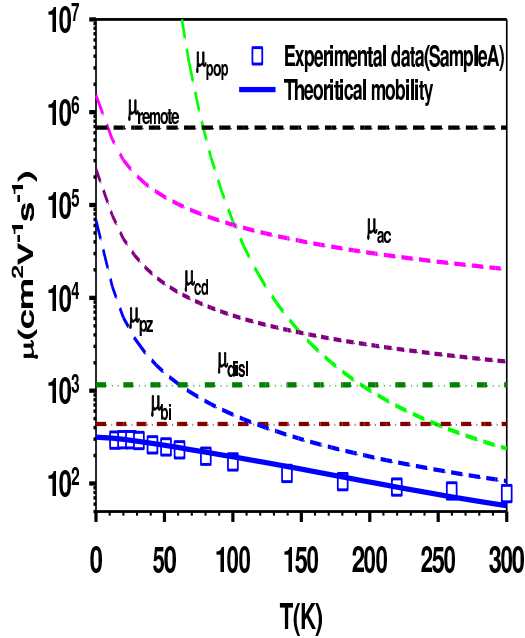}
\caption{\label{pslat5}Experimental and calculated temperature dependence of 
electron mobility curves for Zn$_0.82$Mg$_0.18$/ZnO.}
\end{figure}

\begin{figure}
\includegraphics[width=0.40\textwidth,clip=true]{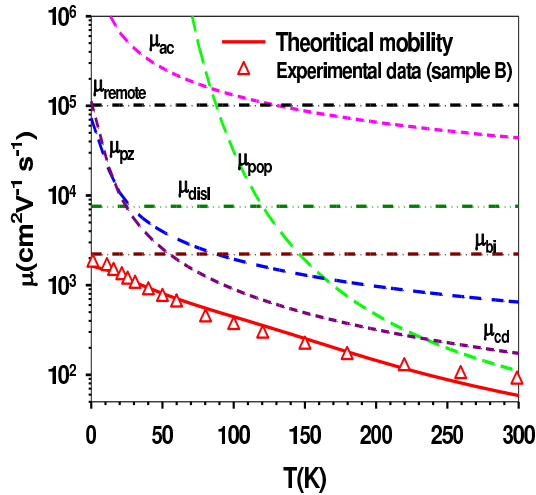}
\caption{\label{pslat6} Experimental and calculated temperature 
dependence of electron mobility curves for Zn$_{0.8}$Mg$_{0.2}$O/ZnO/Zn$_{0.85}$Mg$_{0.15}$O.}
\end{figure}

Fig.~\ref{pslat5} and Fig.~\ref{pslat6} show the temperature dependence of the electron mobility
and the calculated component mobility of the individual scattering process for 
Zn$_{0.82}$Mg$_{0.18}$O (75nm) /ZnO single (sample A) and 
Zn$_{0.8}$Mg$_{0.2}$O(60nm)/ZnO(30nm)/graded-Zn$_{0.85}$Mg$_{0.15}$O(90nm) double (sample B)
heterostructures, respectively.
As can be clearly seen, the electron mobility increases when temperature decreases and
reaches the maximum value of about 290 $cm^{2}V^{-1}s^{-1}$ and 1780 $cm^{2}V^{-1}s^{-1}$ for samples A and B, 
respectively; the difference is considerable. Also, very good consistency is obtained between the 
temperature dependence of the calculated total mobility data and the experimental results.
The fitting parameters are listed in Table~\ref{tab3}.  The investigated structures are with Mg content x
in barrier layer changing from 0.18 to 0.2 and n$_s$ increases from about 1.48$\times$10$^{12}$ and 
1.16$\times10^{14} cm^{-2}$ for samples A and B, respectively. 

\begin{table}[htb]
\caption{\label{tab3} Value of the calculated fitting parameters for the heterostructure samples.}
\begin{ruledtabular}
\begin{tabular}{lcc} 
Fitting parameters &Sample A&Sample B\\
\hline
Dislocation density N$_{disl}$ (m$^{-2}$)                      &7$\times$10$^{12}$        &1$\times$10$^{15}$\\
2D impurity density in the potential                           &                          & \\
well N$_{bi}$ (m$^{-3}$)                                       &9$\times$10$^{23}$        &4.5$\times$10$^{25}$\\
C Parameter                                                    &5.5$\times$10$^{5}$       &9$\times$10$^{5}$\\
\end{tabular}
\end{ruledtabular}
\end{table}

The dislocation scattering and ionized impurity scattering due to interface charges are weakened when 
n$_s$ increases, since the screening effect of the electrons on the scattering centers is improved.

The enhancement of electron mobility in sample B may be associated with the decrease of N$_{disl}$, 
N$_{bi}$ and strain induced fields in this sample. Also, better electron confinement in
heterostructures plays an important role in determining 2DEG mobility via increasing the screening effect 
against ionized impurity and dislocation scattering
~\cite{Weimann,Ando}.

As a result, the produced electric field via piezoelectric polarization charge at the ZnMgO/ZnO 
heterointerface and formation of the conduction band discontinuity at the same interface can lead
to better electron confinement and, then, electron mobility of 
Zn$_{0.8}$Mg$_{0.2}$O/ZnO/Zn$_{0.85}$Mg$_{0.15}$O is enhanced.  

As can be seen in Fig.~\ref{pslat5}:
For sample A, at low temperature ($T \textless 10K$), ionized impurity scattering due to interface
charges and at high temperature ($110\textless T \textless 300 K$) piezoelectric scattering control
electron mobility.
For sample B, at low temperature ($T \textless 60 K$), ionized impurity scattering due to interface 
charges scattering, at medium temperature ($60\textless T \textless 230 K$), crystalline defects
and at high temperature ($T>230K$), polar optical phonon scattering restrict mobility.
It should be noted that the obtained dislocation density matches the accuracy of the 
reported values~\cite{Look9}, in the range 10$^{9}$ to 10$^{11}$ cm$^{-2}$.

\section{Conclusion}
In this paper, the reported experimental data related to the electrical transport properties of 
ZnO/sapphire, Zn$_{0.82}$Mg$_{0.18}$O (75nm) /ZnO single (sample A) and
Zn$_{0.8}$Mg$_{0.2O}$(60nm)/ZnO(30nm)/graded-Zn$_{0.85}$Mg$_{0.15}$O(90nm) double 
(sample B)  heterostructures are quantitatively calculated with different Mg concentrations and 
barrier thickness.
For bulk ZnO, the effect of degenerate layer at the ZnO/sapphire interface on the experimental electron 
concentration and mobility can be removed using a two layer Hall effect model. The fitting 
curve of temperature-dependent corrected electron concentration results in E$_d$= 25meV, 
Na = 1$\times10^{16} cm^{-3}$, and N$_d$=6$\times10^{17} cm^{-3}$. 
The fitting curves of temperature dependence of the corrected electron mobility shows that 
dislocation scattering restricts electron mobility approximately in all temperature ranges,
which is due to large lattice mismatch between ZnO and sapphire. The value of  N$_{disl}$ in this 
sample is obtained about 6.5$\times10^{13} cm^{-2}.$ For Zn$_{0.82}$Mg$_{0.18}$O/ZnO single and 
Zn$_{0.8}$Mg$_{0.2}$O/ZnO/graded-Zn$_{0.85}$Mg$_0.15$O double heterostructures, the temperature
dependence of 2DEG mobility is determined by taking into account all the major scattering mechanisms.
The calculated results for 2DEG  mobility indicate that, in ZnMgO/ZnO/ZnMgO heterostructure,
dislocation and ionized background impurity are effectively suppressed, which is related to
enhancement in the sheet carrier concentration of  2DEG and  better electron confinement in the 
channel because of the produced electric field via piezoelectric polarization charge and formation of
conduction band discontinuity in the heterointerface in this sample.

\bibliography{ref}

\begin{thebibliography}{45}%
\makeatletter
\providecommand \@ifxundefined [1]{%
 \@ifx{#1\undefined}
}%
\providecommand \@ifnum [1]{%
 \ifnum #1\expandafter \@firstoftwo
 \else \expandafter \@secondoftwo
 \fi
}%
\providecommand \@ifx [1]{%
 \ifx #1\expandafter \@firstoftwo
 \else \expandafter \@secondoftwo
 \fi
}%
\providecommand \natexlab [1]{#1}%
\providecommand \enquote  [1]{``#1''}%
\providecommand \bibnamefont  [1]{#1}%
\providecommand \bibfnamefont [1]{#1}%
\providecommand \citenamefont [1]{#1}%
\providecommand \href@noop [0]{\@secondoftwo}%
\providecommand \href [0]{\begingroup \@sanitize@url \@href}%
\providecommand \@href[1]{\@@startlink{#1}\@@href}%
\providecommand \@@href[1]{\endgroup#1\@@endlink}%
\providecommand \@sanitize@url [0]{\catcode `\\12\catcode `\$12\catcode
  `\&12\catcode `\#12\catcode `\^12\catcode `\_12\catcode `\%12\relax}%
\providecommand \@@startlink[1]{}%
\providecommand \@@endlink[0]{}%
\providecommand \url  [0]{\begingroup\@sanitize@url \@url }%
\providecommand \@url [1]{\endgroup\@href {#1}{\urlprefix }}%
\providecommand \urlprefix  [0]{URL }%
\providecommand \Eprint [0]{\href }%
\providecommand \doibase [0]{http://dx.doi.org/}%
\providecommand \selectlanguage [0]{\@gobble}%
\providecommand \bibinfo  [0]{\@secondoftwo}%
\providecommand \bibfield  [0]{\@secondoftwo}%
\providecommand \translation [1]{[#1]}%
\providecommand \BibitemOpen [0]{}%
\providecommand \bibitemStop [0]{}%
\providecommand \bibitemNoStop [0]{.\EOS\space}%
\providecommand \EOS [0]{\spacefactor3000\relax}%
\providecommand \BibitemShut  [1]{\csname bibitem#1\endcsname}%
\let\auto@bib@innerbib\@empty
\bibitem [{\citenamefont {Dai}\ \emph {et~al.}(2011)\citenamefont {Dai},
  \citenamefont {Han}, \citenamefont {Wu}, \citenamefont {Fang}, \citenamefont
  {Xiong}, \citenamefont {Tian}, \citenamefont {Yu}, \citenamefont {He},\ and\
  \citenamefont {Chen}}]{Dai2011}%
  \BibitemOpen
  \bibfield  {author} {\bibinfo {author} {\bibfnamefont {J.}~\bibnamefont
  {Dai}}, \bibinfo {author} {\bibfnamefont {X.}~\bibnamefont {Han}}, \bibinfo
  {author} {\bibfnamefont {Z.}~\bibnamefont {Wu}}, \bibinfo {author}
  {\bibfnamefont {Y.}~\bibnamefont {Fang}}, \bibinfo {author} {\bibfnamefont
  {H.}~\bibnamefont {Xiong}}, \bibinfo {author} {\bibfnamefont
  {Y.}~\bibnamefont {Tian}}, \bibinfo {author} {\bibfnamefont {C.}~\bibnamefont
  {Yu}}, \bibinfo {author} {\bibfnamefont {Q.}~\bibnamefont {He}}, \ and\
  \bibinfo {author} {\bibfnamefont {C.}~\bibnamefont {Chen}},\ }\href {\doibase
  10.1007/s11664-011-1511-6} {\bibfield  {journal} {\bibinfo  {journal}
  {Journal of Electronic Materials}\ }\textbf {\bibinfo {volume} {40}},\
  \bibinfo {pages} {446} (\bibinfo {year} {2011})}\BibitemShut {NoStop}%
\bibitem [{\citenamefont {Meng}\ \emph {et~al.}(2011)\citenamefont {Meng},
  \citenamefont {Zheng}, \citenamefont {Cheng}, \citenamefont {Li},
  \citenamefont {Huang}, \citenamefont {Gu},\ and\ \citenamefont
  {Zhang}}]{Meng}%
  \BibitemOpen
  \bibfield  {author} {\bibinfo {author} {\bibfnamefont {L.}~\bibnamefont
  {Meng}}, \bibinfo {author} {\bibfnamefont {L.}~\bibnamefont {Zheng}},
  \bibinfo {author} {\bibfnamefont {L.}~\bibnamefont {Cheng}}, \bibinfo
  {author} {\bibfnamefont {G.}~\bibnamefont {Li}}, \bibinfo {author}
  {\bibfnamefont {L.}~\bibnamefont {Huang}}, \bibinfo {author} {\bibfnamefont
  {Y.}~\bibnamefont {Gu}}, \ and\ \bibinfo {author} {\bibfnamefont
  {F.}~\bibnamefont {Zhang}},\ }\href@noop {} {\bibfield  {journal} {\bibinfo
  {journal} {J. Mater. Chem.}\ }\textbf {\bibinfo {volume} {21}},\ \bibinfo
  {pages} {11418} (\bibinfo {year} {2011})}\BibitemShut {NoStop}%
\bibitem [{\citenamefont {Park}\ \emph {et~al.}(2013)\citenamefont {Park},
  \citenamefont {Hwang}, \citenamefont {Seo},\ and\ \citenamefont
  {Seo}}]{kpark}%
  \BibitemOpen
  \bibfield  {author} {\bibinfo {author} {\bibfnamefont {K.}~\bibnamefont
  {Park}}, \bibinfo {author} {\bibfnamefont {H.}~\bibnamefont {Hwang}},
  \bibinfo {author} {\bibfnamefont {J.}~\bibnamefont {Seo}}, \ and\ \bibinfo
  {author} {\bibfnamefont {W.-S.}\ \bibnamefont {Seo}},\ }\href {\doibase
  http://dx.doi.org/10.1016/j.energy.2013.03.023} {\bibfield  {journal}
  {\bibinfo  {journal} {Energy}\ }\textbf {\bibinfo {volume} {54}},\ \bibinfo
  {pages} {139 } (\bibinfo {year} {2013})}\BibitemShut {NoStop}%
\bibitem [{\citenamefont {Pholnak}\ \emph {et~al.}(2013)\citenamefont
  {Pholnak}, \citenamefont {Suwanboon},\ and\ \citenamefont
  {Sirisathitkul}}]{Pholnak}%
  \BibitemOpen
  \bibfield  {author} {\bibinfo {author} {\bibfnamefont {C.}~\bibnamefont
  {Pholnak}}, \bibinfo {author} {\bibfnamefont {S.}~\bibnamefont {Suwanboon}},
  \ and\ \bibinfo {author} {\bibfnamefont {C.}~\bibnamefont {Sirisathitkul}},\
  }\href {\doibase 10.1007/s10854-013-1516-4} {\bibfield  {journal} {\bibinfo
  {journal} {Journal of Materials Science: Materials in Electronics}\ }\textbf
  {\bibinfo {volume} {24}},\ \bibinfo {pages} {5014} (\bibinfo {year}
  {2013})}\BibitemShut {NoStop}%
\bibitem [{\citenamefont {H.Morkoc}\ and\ \citenamefont
  {Ozgur}(2009)}]{Morkoc}%
  \BibitemOpen
  \bibfield  {author} {\bibinfo {author} {\bibnamefont {H.Morkoc}}\ and\
  \bibinfo {author} {\bibfnamefont {U.}~\bibnamefont {Ozgur}},\ }\href@noop {}
  {\emph {\bibinfo {title} {Zinc Oxide Fundamentals, Materials and Device
  Technology}}}\ (\bibinfo  {publisher} {WILEY-VCH},\ \bibinfo {year}
  {2009})\BibitemShut {NoStop}%
\bibitem [{\citenamefont {Wang}\ \emph
  {et~al.}(2015{\natexlab{a}})\citenamefont {Wang}, \citenamefont {Bao},
  \citenamefont {Zhao}, \citenamefont {Zhang}, \citenamefont {Dong},\ and\
  \citenamefont {Pan}}]{Wang}%
  \BibitemOpen
  \bibfield  {author} {\bibinfo {author} {\bibfnamefont {C.}~\bibnamefont
  {Wang}}, \bibinfo {author} {\bibfnamefont {R.}~\bibnamefont {Bao}}, \bibinfo
  {author} {\bibfnamefont {K.}~\bibnamefont {Zhao}}, \bibinfo {author}
  {\bibfnamefont {T.}~\bibnamefont {Zhang}}, \bibinfo {author} {\bibfnamefont
  {L.}~\bibnamefont {Dong}}, \ and\ \bibinfo {author} {\bibfnamefont
  {C.}~\bibnamefont {Pan}},\ }\href {\doibase
  http://dx.doi.org/10.1016/j.nanoen.2014.11.033} {\bibfield  {journal}
  {\bibinfo  {journal} {Nano Energy}\ }\textbf {\bibinfo {volume} {14}},\
  \bibinfo {pages} {364 } (\bibinfo {year} {2015}{\natexlab{a}})},\ \bibinfo
  {note} {special issue on the 2nd International Conference on Nanogenerators
  and Piezotronics (NGPT 2014)2nd International Conference on Nanogenerators
  and Piezotronics}\BibitemShut {NoStop}%
\bibitem [{\citenamefont {Wang}\ \emph
  {et~al.}(2015{\natexlab{b}})\citenamefont {Wang}, \citenamefont {Zhao},
  \citenamefont {Wu}, \citenamefont {Dong}, \citenamefont {Zhang},
  \citenamefont {Wu}, \citenamefont {Ma},\ and\ \citenamefont {Du}}]{Wnag2}%
  \BibitemOpen
  \bibfield  {author} {\bibinfo {author} {\bibfnamefont {H.}~\bibnamefont
  {Wang}}, \bibinfo {author} {\bibfnamefont {Y.}~\bibnamefont {Zhao}}, \bibinfo
  {author} {\bibfnamefont {C.}~\bibnamefont {Wu}}, \bibinfo {author}
  {\bibfnamefont {X.}~\bibnamefont {Dong}}, \bibinfo {author} {\bibfnamefont
  {B.}~\bibnamefont {Zhang}}, \bibinfo {author} {\bibfnamefont
  {G.}~\bibnamefont {Wu}}, \bibinfo {author} {\bibfnamefont {Y.}~\bibnamefont
  {Ma}}, \ and\ \bibinfo {author} {\bibfnamefont {G.}~\bibnamefont {Du}},\
  }\href {\doibase http://dx.doi.org/10.1016/j.jlumin.2014.09.007} {\bibfield
  {journal} {\bibinfo  {journal} {Journal of Luminescence}\ }\textbf {\bibinfo
  {volume} {158}},\ \bibinfo {pages} {6 } (\bibinfo {year}
  {2015}{\natexlab{b}})}\BibitemShut {NoStop}%
\bibitem [{\citenamefont {Kwon}\ \emph {et~al.}(2015)\citenamefont {Kwon},
  \citenamefont {Hong}, \citenamefont {Kwon}, \citenamefont {Park},
  \citenamefont {Yoo}, \citenamefont {Kim}, \citenamefont {Grigoropoulos},
  \citenamefont {Oh},\ and\ \citenamefont {Kim}}]{Kwon}%
  \BibitemOpen
  \bibfield  {author} {\bibinfo {author} {\bibfnamefont {J.}~\bibnamefont
  {Kwon}}, \bibinfo {author} {\bibfnamefont {Y.~K.}\ \bibnamefont {Hong}},
  \bibinfo {author} {\bibfnamefont {H.-J.}\ \bibnamefont {Kwon}}, \bibinfo
  {author} {\bibfnamefont {Y.~J.}\ \bibnamefont {Park}}, \bibinfo {author}
  {\bibfnamefont {B.}~\bibnamefont {Yoo}}, \bibinfo {author} {\bibfnamefont
  {J.}~\bibnamefont {Kim}}, \bibinfo {author} {\bibfnamefont {C.~P.}\
  \bibnamefont {Grigoropoulos}}, \bibinfo {author} {\bibfnamefont {M.~S.}\
  \bibnamefont {Oh}}, \ and\ \bibinfo {author} {\bibfnamefont {S.}~\bibnamefont
  {Kim}},\ }\href {http://stacks.iop.org/0957-4484/26/i=3/a=035202} {\bibfield
  {journal} {\bibinfo  {journal} {Nanotechnology}\ }\textbf {\bibinfo {volume}
  {26}},\ \bibinfo {pages} {035202} (\bibinfo {year} {2015})}\BibitemShut
  {NoStop}%
\bibitem [{\citenamefont {Guo}\ \emph {et~al.}(2012)\citenamefont {Guo},
  \citenamefont {Zhang}, \citenamefont {Zhao}, \citenamefont {Li},
  \citenamefont {Zhang}, \citenamefont {Jiang},\ and\ \citenamefont
  {Shen}}]{Guoa}%
  \BibitemOpen
  \bibfield  {author} {\bibinfo {author} {\bibfnamefont {L.}~\bibnamefont
  {Guo}}, \bibinfo {author} {\bibfnamefont {H.}~\bibnamefont {Zhang}}, \bibinfo
  {author} {\bibfnamefont {D.}~\bibnamefont {Zhao}}, \bibinfo {author}
  {\bibfnamefont {B.}~\bibnamefont {Li}}, \bibinfo {author} {\bibfnamefont
  {Z.}~\bibnamefont {Zhang}}, \bibinfo {author} {\bibfnamefont
  {M.}~\bibnamefont {Jiang}}, \ and\ \bibinfo {author} {\bibfnamefont
  {D.}~\bibnamefont {Shen}},\ }\href {\doibase
  http://dx.doi.org/10.1016/j.snb.2011.08.049} {\bibfield  {journal} {\bibinfo
  {journal} {Sensors and Actuators B: Chemical}\ }\textbf {\bibinfo {volume}
  {166–167}},\ \bibinfo {pages} {12 } (\bibinfo {year} {2012})}\BibitemShut
  {NoStop}%
\bibitem [{\citenamefont {Shi}\ \emph {et~al.}(2015)\citenamefont {Shi},
  \citenamefont {Zhang}, \citenamefont {Cui}, \citenamefont {Zhuang},
  \citenamefont {Wu}, \citenamefont {Chu}, \citenamefont {Dong}, \citenamefont
  {Zhang},\ and\ \citenamefont {Du}}]{Shi}%
  \BibitemOpen
  \bibfield  {author} {\bibinfo {author} {\bibfnamefont {Z.-F.}\ \bibnamefont
  {Shi}}, \bibinfo {author} {\bibfnamefont {Y.-T.}\ \bibnamefont {Zhang}},
  \bibinfo {author} {\bibfnamefont {X.-J.}\ \bibnamefont {Cui}}, \bibinfo
  {author} {\bibfnamefont {S.-W.}\ \bibnamefont {Zhuang}}, \bibinfo {author}
  {\bibfnamefont {B.}~\bibnamefont {Wu}}, \bibinfo {author} {\bibfnamefont
  {X.-W.}\ \bibnamefont {Chu}}, \bibinfo {author} {\bibfnamefont
  {X.}~\bibnamefont {Dong}}, \bibinfo {author} {\bibfnamefont {B.-L.}\
  \bibnamefont {Zhang}}, \ and\ \bibinfo {author} {\bibfnamefont {G.-T.}\
  \bibnamefont {Du}},\ }\href@noop {} {\bibfield  {journal} {\bibinfo
  {journal} {Phys. Chem. Chem. Phys.}\ }\textbf {\bibinfo {volume} {17}},\
  \bibinfo {pages} {13813} (\bibinfo {year} {2015})}\BibitemShut {NoStop}%
\bibitem [{\citenamefont {Szymański}\ \emph {et~al.}(2014)\citenamefont
  {Szymański}, \citenamefont {Teisseyre},\ and\ \citenamefont
  {Kozanecki}}]{Szymanski}%
  \BibitemOpen
  \bibfield  {author} {\bibinfo {author} {\bibfnamefont {M.}~\bibnamefont
  {Szymański}}, \bibinfo {author} {\bibfnamefont {H.}~\bibnamefont
  {Teisseyre}}, \ and\ \bibinfo {author} {\bibfnamefont {A.}~\bibnamefont
  {Kozanecki}},\ }\href {\doibase 10.1002/pssa.201300764} {\bibfield  {journal}
  {\bibinfo  {journal} {physica status solidi (a)}\ }\textbf {\bibinfo {volume}
  {211}},\ \bibinfo {pages} {2105} (\bibinfo {year} {2014})}\BibitemShut
  {NoStop}%
\bibitem [{\citenamefont {Bian}\ \emph {et~al.}(2014)\citenamefont {Bian},
  \citenamefont {Miao}, \citenamefont {Qin}, \citenamefont {Zhang},
  \citenamefont {Liu},\ and\ \citenamefont {Liu}}]{Bian}%
  \BibitemOpen
  \bibfield  {author} {\bibinfo {author} {\bibfnamefont {J.}~\bibnamefont
  {Bian}}, \bibinfo {author} {\bibfnamefont {L.}~\bibnamefont {Miao}}, \bibinfo
  {author} {\bibfnamefont {F.}~\bibnamefont {Qin}}, \bibinfo {author}
  {\bibfnamefont {D.}~\bibnamefont {Zhang}}, \bibinfo {author} {\bibfnamefont
  {W.}~\bibnamefont {Liu}}, \ and\ \bibinfo {author} {\bibfnamefont
  {H.}~\bibnamefont {Liu}},\ }\href {\doibase
  http://dx.doi.org/10.1016/j.mssp.2014.04.030} {\bibfield  {journal} {\bibinfo
   {journal} {Materials Science in Semiconductor Processing}\ }\textbf
  {\bibinfo {volume} {26}},\ \bibinfo {pages} {182 } (\bibinfo {year}
  {2014})}\BibitemShut {NoStop}%
\bibitem [{\citenamefont {Sang}\ \emph {et~al.}(2013)\citenamefont {Sang},
  \citenamefont {Yang}, \citenamefont {Liu}, \citenamefont {Zhao},
  \citenamefont {Liu}, \citenamefont {Gu}, \citenamefont {Wei}, \citenamefont
  {Liu}, \citenamefont {Zhu},\ and\ \citenamefont {Wang}}]{Sang}%
  \BibitemOpen
  \bibfield  {author} {\bibinfo {author} {\bibfnamefont {L.}~\bibnamefont
  {Sang}}, \bibinfo {author} {\bibfnamefont {S.~Y.}\ \bibnamefont {Yang}},
  \bibinfo {author} {\bibfnamefont {G.~P.}\ \bibnamefont {Liu}}, \bibinfo
  {author} {\bibfnamefont {G.~J.}\ \bibnamefont {Zhao}}, \bibinfo {author}
  {\bibfnamefont {C.~B.}\ \bibnamefont {Liu}}, \bibinfo {author} {\bibfnamefont
  {C.~Y.}\ \bibnamefont {Gu}}, \bibinfo {author} {\bibfnamefont {H.~Y.}\
  \bibnamefont {Wei}}, \bibinfo {author} {\bibfnamefont {X.~L.}\ \bibnamefont
  {Liu}}, \bibinfo {author} {\bibfnamefont {Q.~S.}\ \bibnamefont {Zhu}}, \ and\
  \bibinfo {author} {\bibfnamefont {Z.~G.}\ \bibnamefont {Wang}},\ }\href
  {\doibase 10.1109/TED.2013.2255599} {\bibfield  {journal} {\bibinfo
  {journal} {Electron Devices, IEEE Transactions on}\ }\textbf {\bibinfo
  {volume} {60}},\ \bibinfo {pages} {2077} (\bibinfo {year}
  {2013})}\BibitemShut {NoStop}%
\bibitem [{\citenamefont {Wang}\ \emph {et~al.}(2013)\citenamefont {Wang},
  \citenamefont {Liao}, \citenamefont {Chueh}, \citenamefont {Lai},
  \citenamefont {Chou},\ and\ \citenamefont {Ting}}]{Wang3}%
  \BibitemOpen
  \bibfield  {author} {\bibinfo {author} {\bibfnamefont {H.-C.}\ \bibnamefont
  {Wang}}, \bibinfo {author} {\bibfnamefont {C.-H.}\ \bibnamefont {Liao}},
  \bibinfo {author} {\bibfnamefont {Y.-L.}\ \bibnamefont {Chueh}}, \bibinfo
  {author} {\bibfnamefont {C.-C.}\ \bibnamefont {Lai}}, \bibinfo {author}
  {\bibfnamefont {P.-C.}\ \bibnamefont {Chou}}, \ and\ \bibinfo {author}
  {\bibfnamefont {S.-Y.}\ \bibnamefont {Ting}},\ }\href@noop {} {\bibfield
  {journal} {\bibinfo  {journal} {Opt. Mater. Express}\ }\textbf {\bibinfo
  {volume} {3}},\ \bibinfo {pages} {295} (\bibinfo {year} {2013})}\BibitemShut
  {NoStop}%
\bibitem [{\citenamefont {Kuznetsov}\ \emph {et~al.}(2010)\citenamefont
  {Kuznetsov}, \citenamefont {Lusanov}, \citenamefont {Yakushcheva},
  \citenamefont {Jitov}, \citenamefont {Zakharov}, \citenamefont
  {Kotelyanskii},\ and\ \citenamefont {Kozlovsky}}]{Kuznetsov}%
  \BibitemOpen
  \bibfield  {author} {\bibinfo {author} {\bibfnamefont {P.}~\bibnamefont
  {Kuznetsov}}, \bibinfo {author} {\bibfnamefont {V.}~\bibnamefont {Lusanov}},
  \bibinfo {author} {\bibfnamefont {G.}~\bibnamefont {Yakushcheva}}, \bibinfo
  {author} {\bibfnamefont {V.}~\bibnamefont {Jitov}}, \bibinfo {author}
  {\bibfnamefont {L.}~\bibnamefont {Zakharov}}, \bibinfo {author}
  {\bibfnamefont {I.}~\bibnamefont {Kotelyanskii}}, \ and\ \bibinfo {author}
  {\bibfnamefont {V.}~\bibnamefont {Kozlovsky}},\ }\href {\doibase
  10.1002/pssc.200983176} {\bibfield  {journal} {\bibinfo  {journal} {physica
  status solidi (c)}\ }\textbf {\bibinfo {volume} {7}},\ \bibinfo {pages}
  {1568} (\bibinfo {year} {2010})}\BibitemShut {NoStop}%
\bibitem [{\citenamefont {Pedro~Barquinha}(2006)}]{Barquinha}%
  \BibitemOpen
  \bibfield  {author} {\bibinfo {author} {\bibfnamefont {A.~G. A. P. A. M. L.
  P. R.~M.}\ \bibnamefont {Pedro~Barquinha}, \bibfnamefont
  {Elvira~Fortunato}},\ }\href@noop {} {\bibfield  {journal} {\bibinfo
  {journal} {Materials Science Forum}\ }\textbf {\bibinfo {volume} {514-516}},\
  \bibinfo {pages} {68} (\bibinfo {year} {2006})}\BibitemShut {NoStop}%
\bibitem [{\citenamefont {AMIRABBASI}(2013)}]{Amirabbasi}%
  \BibitemOpen
  \bibfield  {author} {\bibinfo {author} {\bibfnamefont {M.}~\bibnamefont
  {AMIRABBASI}},\ }\href {\doibase 10.1142/S0217984913501704} {\bibfield
  {journal} {\bibinfo  {journal} {Modern Physics Letters B}\ }\textbf {\bibinfo
  {volume} {27}},\ \bibinfo {pages} {1350170} (\bibinfo {year}
  {2013})}\BibitemShut {NoStop}%
\bibitem [{\citenamefont {Ji}\ \emph {et~al.}(2014)\citenamefont {Ji},
  \citenamefont {Zhu}, \citenamefont {Chen}, \citenamefont {Su}, \citenamefont
  {Chen}, \citenamefont {Gui}, \citenamefont {Xiang},\ and\ \citenamefont
  {Tang}}]{Ji}%
  \BibitemOpen
  \bibfield  {author} {\bibinfo {author} {\bibfnamefont {X.}~\bibnamefont
  {Ji}}, \bibinfo {author} {\bibfnamefont {Y.}~\bibnamefont {Zhu}}, \bibinfo
  {author} {\bibfnamefont {M.}~\bibnamefont {Chen}}, \bibinfo {author}
  {\bibfnamefont {L.}~\bibnamefont {Su}}, \bibinfo {author} {\bibfnamefont
  {A.}~\bibnamefont {Chen}}, \bibinfo {author} {\bibfnamefont {X.}~\bibnamefont
  {Gui}}, \bibinfo {author} {\bibfnamefont {R.}~\bibnamefont {Xiang}}, \ and\
  \bibinfo {author} {\bibfnamefont {Z.}~\bibnamefont {Tang}},\ }\href
  {http://dx.doi.org/10.1038/srep04185} {\bibfield  {journal} {\bibinfo
  {journal} {Scientific Reports}\ }\textbf {\bibinfo {volume} {4}},\ \bibinfo
  {pages} {4185 EP } (\bibinfo {year} {2014})},\ \bibinfo {note}
  {article}\BibitemShut {NoStop}%
\bibitem [{\citenamefont {Meng}(2015)}]{Meng3}%
  \BibitemOpen
  \bibfield  {author} {\bibinfo {author} {\bibfnamefont {J.~L. Q. H.~X.}\
  \bibnamefont {Meng}, \bibfnamefont {Li~Zhang}},\ }\href
  {http://dx.doi.org/10.1155/2015/694234} {\bibfield  {journal} {\bibinfo
  {journal} {Journal of Nanomaterials}\ ,\ \bibinfo {pages} {694234}} (\bibinfo
  {year} {2015})},\ \bibinfo {note} {article}\BibitemShut {NoStop}%
\bibitem [{\citenamefont {Ye}\ \emph {et~al.}(2012)\citenamefont {Ye},
  \citenamefont {Ter~Lim}, \citenamefont {Bosman}, \citenamefont {Gu},
  \citenamefont {Zheng}, \citenamefont {Tan}, \citenamefont {Jagadish},
  \citenamefont {Sun},\ and\ \citenamefont {Teo}}]{Ye}%
  \BibitemOpen
  \bibfield  {author} {\bibinfo {author} {\bibfnamefont {J.}~\bibnamefont
  {Ye}}, \bibinfo {author} {\bibfnamefont {S.}~\bibnamefont {Ter~Lim}},
  \bibinfo {author} {\bibfnamefont {M.}~\bibnamefont {Bosman}}, \bibinfo
  {author} {\bibfnamefont {S.}~\bibnamefont {Gu}}, \bibinfo {author}
  {\bibfnamefont {Y.}~\bibnamefont {Zheng}}, \bibinfo {author} {\bibfnamefont
  {H.~H.}\ \bibnamefont {Tan}}, \bibinfo {author} {\bibfnamefont
  {C.}~\bibnamefont {Jagadish}}, \bibinfo {author} {\bibfnamefont
  {X.}~\bibnamefont {Sun}}, \ and\ \bibinfo {author} {\bibfnamefont {K.~L.}\
  \bibnamefont {Teo}},\ }\href {http://dx.doi.org/10.1038/srep00533} {\bibfield
   {journal} {\bibinfo  {journal} {Scientific Reports}\ }\textbf {\bibinfo
  {volume} {2}},\ \bibinfo {pages} {533 EP } (\bibinfo {year} {2012})},\
  \bibinfo {note} {article}\BibitemShut {NoStop}%
\bibitem [{\citenamefont {D.C.Look}(1998)}]{Look}%
  \BibitemOpen
  \bibfield  {author} {\bibinfo {author} {\bibnamefont {D.C.Look}},\
  }\href@noop {} {\emph {\bibinfo {title} {Electrical characterization of GaAs
  Material and devices}}}\ (\bibinfo  {publisher} {WILEY-VCH},\ \bibinfo {year}
  {1998})\BibitemShut {NoStop}%
\bibitem [{\citenamefont {Furno}\ \emph {et~al.}(2008)\citenamefont {Furno},
  \citenamefont {Bertazzi}, \citenamefont {Goano}, \citenamefont {Ghione},\
  and\ \citenamefont {Bellotti}}]{Furno}%
  \BibitemOpen
  \bibfield  {author} {\bibinfo {author} {\bibfnamefont {E.}~\bibnamefont
  {Furno}}, \bibinfo {author} {\bibfnamefont {F.}~\bibnamefont {Bertazzi}},
  \bibinfo {author} {\bibfnamefont {M.}~\bibnamefont {Goano}}, \bibinfo
  {author} {\bibfnamefont {G.}~\bibnamefont {Ghione}}, \ and\ \bibinfo {author}
  {\bibfnamefont {E.}~\bibnamefont {Bellotti}},\ }\href {\doibase
  http://dx.doi.org/10.1016/j.sse.2008.08.001} {\bibfield  {journal} {\bibinfo
  {journal} {Solid-State Electronics}\ }\textbf {\bibinfo {volume} {52}},\
  \bibinfo {pages} {1796 } (\bibinfo {year} {2008})}\BibitemShut {NoStop}%
\bibitem [{\citenamefont {Rode}(1975)}]{Rode}%
  \BibitemOpen
  \bibfield  {author} {\bibinfo {author} {\bibfnamefont {D.}~\bibnamefont
  {Rode}}\ }(\bibinfo  {publisher} {Elsevier},\ \bibinfo {year} {1975})\ pp.\
  \bibinfo {pages} {1 -- 89}\BibitemShut {NoStop}%
\bibitem [{\citenamefont {Look}\ \emph {et~al.}(1998)\citenamefont {Look},
  \citenamefont {Reynolds}, \citenamefont {Sizelove}, \citenamefont {Jones},
  \citenamefont {Litton}, \citenamefont {Cantwell},\ and\ \citenamefont
  {Harsch}}]{Look2}%
  \BibitemOpen
  \bibfield  {author} {\bibinfo {author} {\bibfnamefont {D.}~\bibnamefont
  {Look}}, \bibinfo {author} {\bibfnamefont {D.}~\bibnamefont {Reynolds}},
  \bibinfo {author} {\bibfnamefont {J.}~\bibnamefont {Sizelove}}, \bibinfo
  {author} {\bibfnamefont {R.}~\bibnamefont {Jones}}, \bibinfo {author}
  {\bibfnamefont {C.}~\bibnamefont {Litton}}, \bibinfo {author} {\bibfnamefont
  {G.}~\bibnamefont {Cantwell}}, \ and\ \bibinfo {author} {\bibfnamefont
  {W.}~\bibnamefont {Harsch}},\ }\href {\doibase
  http://dx.doi.org/10.1016/S0038-1098(97)10145-4} {\bibfield  {journal}
  {\bibinfo  {journal} {Solid State Communications}\ }\textbf {\bibinfo
  {volume} {105}},\ \bibinfo {pages} {399 } (\bibinfo {year}
  {1998})}\BibitemShut {NoStop}%
\bibitem [{\citenamefont {Anderson}\ and\ \citenamefont
  {Apsley}(1986)}]{Anderson}%
  \BibitemOpen
  \bibfield  {author} {\bibinfo {author} {\bibfnamefont {D.~A.}\ \bibnamefont
  {Anderson}}\ and\ \bibinfo {author} {\bibfnamefont {N.}~\bibnamefont
  {Apsley}},\ }\href {http://stacks.iop.org/0268-1242/1/i=3/a=006} {\bibfield
  {journal} {\bibinfo  {journal} {Semiconductor Science and Technology}\
  }\textbf {\bibinfo {volume} {1}},\ \bibinfo {pages} {187} (\bibinfo {year}
  {1986})}\BibitemShut {NoStop}%
\bibitem [{\citenamefont {Ehrenreich}(1959)}]{Ehrenreich}%
  \BibitemOpen
  \bibfield  {author} {\bibinfo {author} {\bibfnamefont {H.}~\bibnamefont
  {Ehrenreich}},\ }\href {\doibase
  http://dx.doi.org/10.1016/0022-3697(59)90297-5} {\bibfield  {journal}
  {\bibinfo  {journal} {Journal of Physics and Chemistry of Solids}\ }\textbf
  {\bibinfo {volume} {8}},\ \bibinfo {pages} {130 } (\bibinfo {year}
  {1959})}\BibitemShut {NoStop}%
\bibitem [{\citenamefont {Chattopadhyay}\ and\ \citenamefont
  {Queisser}(1981)}]{Brooks}%
  \BibitemOpen
  \bibfield  {author} {\bibinfo {author} {\bibfnamefont {D.}~\bibnamefont
  {Chattopadhyay}}\ and\ \bibinfo {author} {\bibfnamefont {H.~J.}\ \bibnamefont
  {Queisser}},\ }\href {\doibase 10.1103/RevModPhys.53.745} {\bibfield
  {journal} {\bibinfo  {journal} {Rev. Mod. Phys.}\ }\textbf {\bibinfo {volume}
  {53}},\ \bibinfo {pages} {745} (\bibinfo {year} {1981})}\BibitemShut
  {NoStop}%
\bibitem [{\citenamefont {Tang}\ \emph {et~al.}(1998)\citenamefont {Tang},
  \citenamefont {Kim}, \citenamefont {Botchkarev}, \citenamefont {Popovici},
  \citenamefont {Hamdani},\ and\ \citenamefont {Morkoç}}]{Tang}%
  \BibitemOpen
  \bibfield  {author} {\bibinfo {author} {\bibfnamefont {H.}~\bibnamefont
  {Tang}}, \bibinfo {author} {\bibfnamefont {W.}~\bibnamefont {Kim}}, \bibinfo
  {author} {\bibfnamefont {A.}~\bibnamefont {Botchkarev}}, \bibinfo {author}
  {\bibfnamefont {G.}~\bibnamefont {Popovici}}, \bibinfo {author}
  {\bibfnamefont {F.}~\bibnamefont {Hamdani}}, \ and\ \bibinfo {author}
  {\bibfnamefont {H.}~\bibnamefont {Morkoç}},\ }\href {\doibase
  http://dx.doi.org/10.1016/S0038-1101(98)00087-2} {\bibfield  {journal}
  {\bibinfo  {journal} {Solid-State Electronics}\ }\textbf {\bibinfo {volume}
  {42}},\ \bibinfo {pages} {839 } (\bibinfo {year} {1998})}\BibitemShut
  {NoStop}%
\bibitem [{\citenamefont {Pödör}(1966)}]{Podor}%
  \BibitemOpen
  \bibfield  {author} {\bibinfo {author} {\bibfnamefont {B.}~\bibnamefont
  {Pödör}},\ }\href {\doibase 10.1002/pssb.19660160264} {\bibfield  {journal}
  {\bibinfo  {journal} {physica status solidi (b)}\ }\textbf {\bibinfo {volume}
  {16}},\ \bibinfo {pages} {K167} (\bibinfo {year} {1966})}\BibitemShut
  {NoStop}%
\bibitem [{\citenamefont {Look}\ and\ \citenamefont {Molnar}(1997)}]{Look5}%
  \BibitemOpen
  \bibfield  {author} {\bibinfo {author} {\bibfnamefont {D.~C.}\ \bibnamefont
  {Look}}\ and\ \bibinfo {author} {\bibfnamefont {R.~J.}\ \bibnamefont
  {Molnar}},\ }\href@noop {} {\bibfield  {journal} {\bibinfo  {journal}
  {Applied Physics Letters}\ }\textbf {\bibinfo {volume} {70}} (\bibinfo {year}
  {1997})}\BibitemShut {NoStop}%
\bibitem [{\citenamefont {Ridley}(1982)}]{Ridley}%
  \BibitemOpen
  \bibfield  {author} {\bibinfo {author} {\bibfnamefont {B.~K.}\ \bibnamefont
  {Ridley}},\ }\href {http://stacks.iop.org/0022-3719/15/i=28/a=021} {\bibfield
   {journal} {\bibinfo  {journal} {Journal of Physics C: Solid State Physics}\
  }\textbf {\bibinfo {volume} {15}},\ \bibinfo {pages} {5899} (\bibinfo {year}
  {1982})}\BibitemShut {NoStop}%
\bibitem [{\citenamefont {Hirakawa}\ and\ \citenamefont
  {Sakaki}(1986)}]{Hirakawa}%
  \BibitemOpen
  \bibfield  {author} {\bibinfo {author} {\bibfnamefont {K.}~\bibnamefont
  {Hirakawa}}\ and\ \bibinfo {author} {\bibfnamefont {H.}~\bibnamefont
  {Sakaki}},\ }\href {\doibase 10.1103/PhysRevB.33.8291} {\bibfield  {journal}
  {\bibinfo  {journal} {Phys. Rev. B}\ }\textbf {\bibinfo {volume} {33}},\
  \bibinfo {pages} {8291} (\bibinfo {year} {1986})}\BibitemShut {NoStop}%
\bibitem [{\citenamefont {Lee}\ \emph {et~al.}(1983)\citenamefont {Lee},
  \citenamefont {Shur}, \citenamefont {Drummond},\ and\ \citenamefont
  {Morkoc}}]{Lee}%
  \BibitemOpen
  \bibfield  {author} {\bibinfo {author} {\bibfnamefont {K.}~\bibnamefont
  {Lee}}, \bibinfo {author} {\bibfnamefont {M.}~\bibnamefont {Shur}}, \bibinfo
  {author} {\bibfnamefont {T.}~\bibnamefont {Drummond}}, \ and\ \bibinfo
  {author} {\bibfnamefont {H.}~\bibnamefont {Morkoc}},\ }\href@noop {}
  {\bibfield  {journal} {\bibinfo  {journal} {Journal of applied physics}\
  }\textbf {\bibinfo {volume} {54}},\ \bibinfo {pages} {6432} (\bibinfo {year}
  {1983})}\BibitemShut {NoStop}%
\bibitem [{\citenamefont {Basu}\ and\ \citenamefont {Nag}(1980)}]{Basu}%
  \BibitemOpen
  \bibfield  {author} {\bibinfo {author} {\bibfnamefont {P.~K.}\ \bibnamefont
  {Basu}}\ and\ \bibinfo {author} {\bibfnamefont {B.~R.}\ \bibnamefont {Nag}},\
  }\href {\doibase 10.1103/PhysRevB.22.4849} {\bibfield  {journal} {\bibinfo
  {journal} {Phys. Rev. B}\ }\textbf {\bibinfo {volume} {22}},\ \bibinfo
  {pages} {4849} (\bibinfo {year} {1980})}\BibitemShut {NoStop}%
\bibitem [{\citenamefont {Price}(1981{\natexlab{a}})}]{price}%
  \BibitemOpen
  \bibfield  {author} {\bibinfo {author} {\bibfnamefont {P.~J.}\ \bibnamefont
  {Price}},\ }\href {\doibase http://dx.doi.org/10.1016/0003-4916(81)90250-5}
  {\bibfield  {journal} {\bibinfo  {journal} {Annals of Physics}\ }\textbf
  {\bibinfo {volume} {133}},\ \bibinfo {pages} {217 } (\bibinfo {year}
  {1981}{\natexlab{a}})}\BibitemShut {NoStop}%
\bibitem [{\citenamefont {Price}(1981{\natexlab{b}})}]{price1981}%
  \BibitemOpen
  \bibfield  {author} {\bibinfo {author} {\bibfnamefont {P.~J.}\ \bibnamefont
  {Price}},\ }\href {\doibase http://dx.doi.org/10.1116/1.571137} {\bibfield
  {journal} {\bibinfo  {journal} {Journal of Vacuum Science and Technology}\
  }\textbf {\bibinfo {volume} {19}},\ \bibinfo {pages} {599} (\bibinfo {year}
  {1981}{\natexlab{b}})}\BibitemShut {NoStop}%
\bibitem [{\citenamefont {Hess}(1979)}]{Hess}%
  \BibitemOpen
  \bibfield  {author} {\bibinfo {author} {\bibfnamefont {K.}~\bibnamefont
  {Hess}},\ }\href@noop {} {\bibfield  {journal} {\bibinfo  {journal} {Applied
  Physics Letters}\ }\textbf {\bibinfo {volume} {35}} (\bibinfo {year}
  {1979})}\BibitemShut {NoStop}%
\bibitem [{\citenamefont {Sah}\ \emph {et~al.}(1972)\citenamefont {Sah},
  \citenamefont {Ning},\ and\ \citenamefont {Tschopp}}]{Sah1972}%
  \BibitemOpen
  \bibfield  {author} {\bibinfo {author} {\bibfnamefont {C.}~\bibnamefont
  {Sah}}, \bibinfo {author} {\bibfnamefont {T.}~\bibnamefont {Ning}}, \ and\
  \bibinfo {author} {\bibfnamefont {L.}~\bibnamefont {Tschopp}},\ }\href
  {\doibase http://dx.doi.org/10.1016/0039-6028(72)90183-5} {\bibfield
  {journal} {\bibinfo  {journal} {Surface Science}\ }\textbf {\bibinfo {volume}
  {32}},\ \bibinfo {pages} {561 } (\bibinfo {year} {1972})}\BibitemShut
  {NoStop}%
\bibitem [{\citenamefont {Das~Sarma}\ and\ \citenamefont
  {Stern}(1985)}]{Sarma}%
  \BibitemOpen
  \bibfield  {author} {\bibinfo {author} {\bibfnamefont {S.}~\bibnamefont
  {Das~Sarma}}\ and\ \bibinfo {author} {\bibfnamefont {F.}~\bibnamefont
  {Stern}},\ }\href {\doibase 10.1103/PhysRevB.32.8442} {\bibfield  {journal}
  {\bibinfo  {journal} {Phys. Rev. B}\ }\textbf {\bibinfo {volume} {32}},\
  \bibinfo {pages} {8442} (\bibinfo {year} {1985})}\BibitemShut {NoStop}%
\bibitem [{\citenamefont {Davies}(1998)}]{Davies}%
  \BibitemOpen
  \bibfield  {author} {\bibinfo {author} {\bibfnamefont {J.~H.}\ \bibnamefont
  {Davies}},\ }\href@noop {} {\emph {\bibinfo {title} {the Physics of Low
  Dimensional Semiconductors}}}\ (\bibinfo  {publisher} {Cambridge University
  Press},\ \bibinfo {year} {1998})\BibitemShut {NoStop}%
\bibitem [{\citenamefont {Look}\ \emph {et~al.}(2003)\citenamefont {Look},
  \citenamefont {Jones}, \citenamefont {Sizelove}, \citenamefont {Garces},
  \citenamefont {Giles},\ and\ \citenamefont {Halliburton}}]{Look9}%
  \BibitemOpen
  \bibfield  {author} {\bibinfo {author} {\bibfnamefont {D.~C.}\ \bibnamefont
  {Look}}, \bibinfo {author} {\bibfnamefont {R.~L.}\ \bibnamefont {Jones}},
  \bibinfo {author} {\bibfnamefont {J.~R.}\ \bibnamefont {Sizelove}}, \bibinfo
  {author} {\bibfnamefont {N.~Y.}\ \bibnamefont {Garces}}, \bibinfo {author}
  {\bibfnamefont {N.~C.}\ \bibnamefont {Giles}}, \ and\ \bibinfo {author}
  {\bibfnamefont {L.~E.}\ \bibnamefont {Halliburton}},\ }\href {\doibase
  10.1002/pssa.200306274} {\bibfield  {journal} {\bibinfo  {journal} {physica
  status solidi (a)}\ }\textbf {\bibinfo {volume} {195}},\ \bibinfo {pages}
  {171} (\bibinfo {year} {2003})}\BibitemShut {NoStop}%
\bibitem [{\citenamefont {Vigué}\ \emph {et~al.}(2001)\citenamefont {Vigué},
  \citenamefont {Vennéguès}, \citenamefont {Deparis}, \citenamefont
  {Vézian}, \citenamefont {Laügt},\ and\ \citenamefont {Faurie}}]{Vigue}%
  \BibitemOpen
  \bibfield  {author} {\bibinfo {author} {\bibfnamefont {F.}~\bibnamefont
  {Vigué}}, \bibinfo {author} {\bibfnamefont {P.}~\bibnamefont {Vennéguès}},
  \bibinfo {author} {\bibfnamefont {C.}~\bibnamefont {Deparis}}, \bibinfo
  {author} {\bibfnamefont {S.}~\bibnamefont {Vézian}}, \bibinfo {author}
  {\bibfnamefont {M.}~\bibnamefont {Laügt}}, \ and\ \bibinfo {author}
  {\bibfnamefont {J.-P.}\ \bibnamefont {Faurie}},\ }\href {\doibase
  http://dx.doi.org/10.1063/1.1412572} {\bibfield  {journal} {\bibinfo
  {journal} {Journal of Applied Physics}\ }\textbf {\bibinfo {volume} {90}},\
  \bibinfo {pages} {5115} (\bibinfo {year} {2001})}\BibitemShut {NoStop}%
\bibitem [{\citenamefont {Baghani}\ and\ \citenamefont
  {OLeary}(2013)}]{Baghani}%
  \BibitemOpen
  \bibfield  {author} {\bibinfo {author} {\bibfnamefont {E.}~\bibnamefont
  {Baghani}}\ and\ \bibinfo {author} {\bibfnamefont {S.~K.}\ \bibnamefont
  {OLeary}},\ }\href {\doibase http://dx.doi.org/10.1063/1.4812492} {\bibfield
  {journal} {\bibinfo  {journal} {Journal of Applied Physics}\ }\textbf
  {\bibinfo {volume} {114}},\ \bibinfo {eid} {023703} (\bibinfo {year}
  {2013}),\ http://dx.doi.org/10.1063/1.4812492}\BibitemShut {NoStop}%
\bibitem [{\citenamefont {Weimann}\ \emph {et~al.}(1998)\citenamefont
  {Weimann}, \citenamefont {Eastman}, \citenamefont {Doppalapudi},
  \citenamefont {Ng},\ and\ \citenamefont {Moustakas}}]{Weimann}%
  \BibitemOpen
  \bibfield  {author} {\bibinfo {author} {\bibfnamefont {N.~G.}\ \bibnamefont
  {Weimann}}, \bibinfo {author} {\bibfnamefont {L.~F.}\ \bibnamefont
  {Eastman}}, \bibinfo {author} {\bibfnamefont {D.}~\bibnamefont
  {Doppalapudi}}, \bibinfo {author} {\bibfnamefont {H.~M.}\ \bibnamefont {Ng}},
  \ and\ \bibinfo {author} {\bibfnamefont {T.~D.}\ \bibnamefont {Moustakas}},\
  }\href@noop {} {\bibfield  {journal} {\bibinfo  {journal} {Journal of Applied
  Physics}\ }\textbf {\bibinfo {volume} {83}} (\bibinfo {year}
  {1998})}\BibitemShut {NoStop}%
\bibitem [{\citenamefont {Ando}\ \emph {et~al.}(1982)\citenamefont {Ando},
  \citenamefont {Fowler},\ and\ \citenamefont {Stern}}]{Ando}%
  \BibitemOpen
  \bibfield  {author} {\bibinfo {author} {\bibfnamefont {T.}~\bibnamefont
  {Ando}}, \bibinfo {author} {\bibfnamefont {A.~B.}\ \bibnamefont {Fowler}}, \
  and\ \bibinfo {author} {\bibfnamefont {F.}~\bibnamefont {Stern}},\ }\href
  {\doibase 10.1103/RevModPhys.54.437} {\bibfield  {journal} {\bibinfo
  {journal} {Rev. Mod. Phys.}\ }\textbf {\bibinfo {volume} {54}},\ \bibinfo
  {pages} {437} (\bibinfo {year} {1982})}\BibitemShut {NoStop}%
\end{thebibliography}%
\end{document}